# Evaluating the Participation of Proactive Scouting Agents in Real-time Human Team Meetings

Hans Schumann
Unanimous AI
San Francisco, CA
Hans@unanimous.ai

Miles Bader
Unanimous AI
Palo Alto, CA
Miles@unanimous.ai

Louis Rosenberg
Unanimous AI
Pismo Beach, CA
ORCID: 0000-0002-2170-7414

Ganesh Mani
Carnegie Mellon University
Pittsburgh, PA
ganeshm@andrew.cmu.edu

**Abstract— An open research question is whether the addition of conversational AI agents to real-time discussions among human teams enhances group performance. This study tested the value of adding proactive scouting agents to teams engaged in collaborative forecasting tasks. These agents were designed to (i) monitor the groupwise discussion in real-time, (ii) identify critical information gaps in the group deliberation, (iii) autonomously find and retrieve relevant information, and (iv) express the acquired content into the conversation using emulated groupwise human etiquette. In a controlled set of trials, small human teams of 4 or 5 members were tasked with collaboratively predicting the outcome of 34 Major League Baseball (MLB) games, either by forecasting the winning team or by predicting the total number of runs scored. In half the trials, the groups were assisted by proactive scouting agents. The other half were human-only deliberations. When tasked with predicting the winning team, the addition of proactive scouting agents to the group discussion significantly increased forecasting accuracy, reducing the predictive mean-absolute error by 25% (p=0.006) and improving raw predictive accuracy from 60.0% to 76.4% (p=0.050). When predicting the total number of runs scored across both teams, the addition of proactive scouting agents yielded no statistical change in group performance. It should be noted that the human participants in this study were significantly worse at predicting the number of runs scored. This may have led to random outcomes in those forecasting trials. Still, the results of this study demonstrate that deploying proactive scouting agents can significantly boost forecasting accuracy during real-time team deliberations when participants possess baseline forecasting skill.**

***Keywords— hyper-communication, conversational forecasting, AI agents, collective intelligence, conversational swarm intelligence, proactive agents, Hyperchat AI, human-AI teaming, co-agents***

## I. Introduction

It is well established that team-based forecasting among human stakeholders outperforms individual forecasters in many traditional business tasks [1]. The accuracy amplification can be attributed to various sources, including the "wisdom of crowds" effect in which aggregating independent inputs across members reduces estimation errors [2]. In addition, deliberation among cross-functional teams has been shown to increase accuracy by synthesizing insights across diverse disciplines [3]. While many studies have been conducted among human forecasting teams of various sizes and compositions, few research efforts have tested multi-human teams supported by conversational AI agents. The present study explores the use of proactive conversational "scouting agents" deployed among human forecasting teams.

As context, researchers proposed a new type of AI agent in 2023 called a "Conversational Contributor Agent" (or co-agent) that supports human teams engaged in real-time discussions [4, 5]. These agents are intended to participate conversationally during group meetings in which a human team works to collaboratively solve problems, make assessments, generate forecasts, or reach decisions [6]. To participate effectively in groupwise human meetings, co-agents are designed to track ongoing conversations in real-time and selectively provide relevant, missing pieces of information (e.g., supporting or opposing evidence or context) that are likely to bolster the unfolding deliberation. The objective of each co-agent is to ensure that a human team is well-informed about the issues under discussion.

As defined, co-agents can be reactive or proactive, meaning they can (i) exclusively respond to a direct request for information from one or more members of the team or, (ii) autonomously identify knowledge gaps in the ongoing group discussion, and search for responsive information without being asked. Co-agents that proactively search for content are sometimes referred to as *scout agents* or *scouting agents* [4-7]. In addition, researchers have proposed proactive co-agents that perform other groupwise collaborative functions, for example, helping a human team brainstorm by contributing creative ideas during a real-time conversation [8,9]. This leads to the example taxonomy for Contributor Agents as shown in Fig. 1 below.

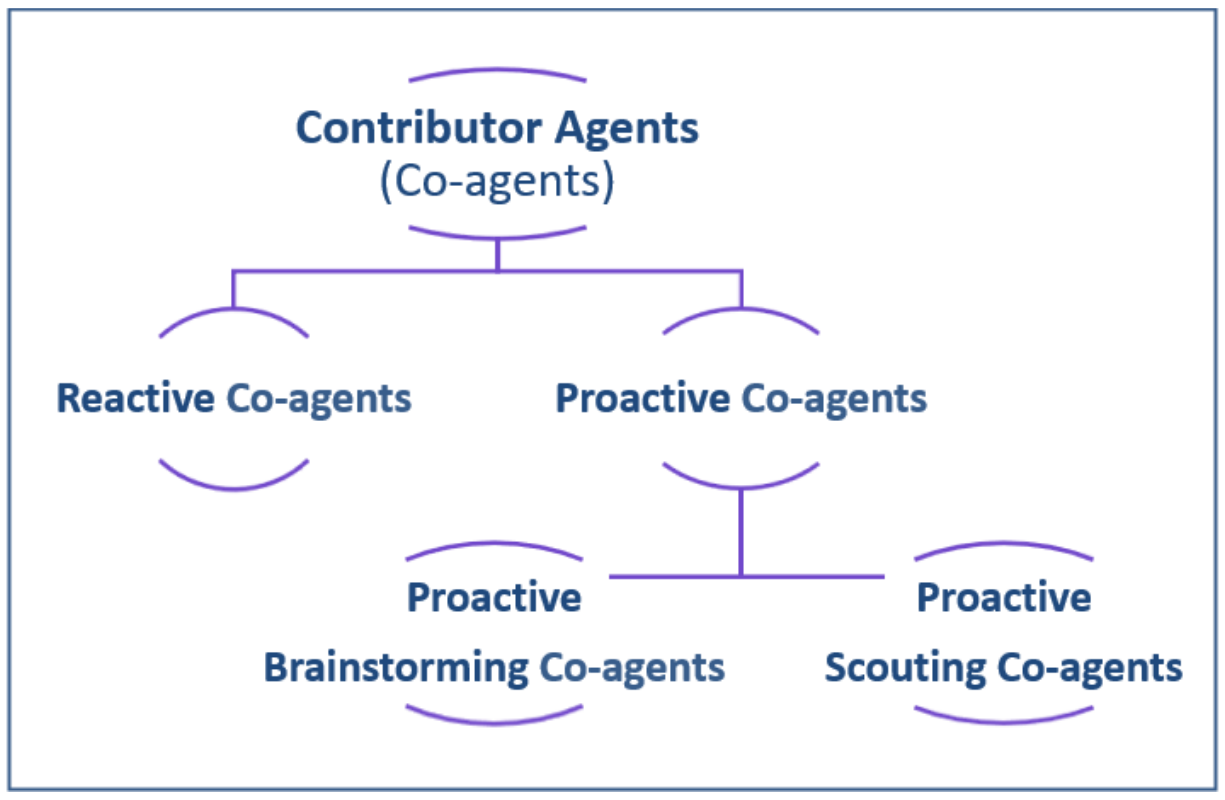

Figure 1: Taxonomy of Contributor Agents (Co-agents)

The first study of interactive AI agents designed to support conversational human teams during group forecasting tasks was published in 2024 [10]. In these experiments, conducted by researchers at Carnegie Mellon University and Unanimous AI, small groups of human participants were given a sports-based forecasting task in which they had to collaboratively select Major League Baseball (MLB) players for a team roster in a real

Fantasy Baseball contest through conversational discussion. A reactive co-agent referred to in that study as an "Infobot" was included in half of the experimental trials. It was pre-loaded with up-to-date baseball statistics that human participants could query by simply asking the Infobot a question. In this way, each small group could discuss and debate their Fantasy Team roster and any question that a team member asked to the Infobot would be part of the ongoing conversation, so all members were informed at the same time, providing a common context [10].

The results of the study suggested a positive impact when adding reactive co-agents to a group deliberation. For example, the study found that the inclusion of reactive co-agents resulted in more balanced discussions among the human team members (i.e., it was less likely for a small subset of participants to dominate the dialog). Statistically, this was measured as significantly lower variance in member contribution ($p=0.039$) when Infobots were present. In addition, post-session surveys indicated that 85% of participants agreed that the decisions their group reached were stronger due to the information provided by the conversational Infobots (with only 4% disagreeing).

Although encouraging, the above study had several noted limitations. First and foremost, it only explored the use of *reactive* co-agents (i.e., contributor agents that waited for one or more participants to ask for information before providing any responsive content). In addition, these agents had access to only a pre-loaded store of curated MLB statistics, which limited their ability to answer questions during groupwise discussions. To better assess the potential of participatory agents in real-time human meetings, it is important to evaluate *proactive* co-agents that actively track the human discussion in real-time, recognize significant knowledge gaps within the groupwise discourse, and proactively search for information that fill those gaps. We refer to these as "proactive scouting co-agents" because they are (i) able to jump into the group discussion without being asked and (ii) can actively scout the open internet for responsive content.

In addition, the prior study did not sufficiently assess the impact of co-agents on forecasting accuracy. To address these shortcomings, a new study was designed and conducted, as detailed in the following sections. In this study, small groups of human sports fans were tasked with collaboratively predicting the outcome of Major League Baseball games through real-time conversational deliberation. We compare our treatment, small groups containing a proactive co-agent, against the control of human-only groups of equivalent size, randomly assigning the treatment to half of the groups in each forecasting question.

## II. PROACTIVE CONTRIBUTOR AGENTS

Proactive contributor agents are designed to identify and fill informational and logical gaps during group deliberations by sourcing and expressing facts, reasoning, and evidence related to the issues being discussed without requiring any human team member to explicitly ask for assistance. By autonomously providing relevant information as natural dialog within an ongoing group discussion, the human participants neither operate without that information nor are forced to break the conversational flow to search for it themselves. To enable these benefits, proactive scouting co-agent must behave as an informationally savvy interlocutor that monitors the dialog and infers gaps in reasoning, missing information, weak evidence, or incorrect claims, and anticipates needs before they are asked.

Architecturally, each local room hosts a dedicated co-agent that repeatedly ingests updated blocks of groupwise dialog from the conversation it is monitoring. A dispatch loop wakes on new activity; an LLM triage stage then scores the recent exchange for informational urgency and research need—asking, in effect, whether the group has voiced a specific factual uncertainty, made a claim that conflicts with known evidence, left a gap that additional information would help fill, and so on. Higher urgency is reserved for targeted lookups (a concrete uncertainty or a direct ask of the agent); while ordinary topic enrichment, such as facts that deepen understanding of the issues being discussed or options being weighed without answering a pinpointed ask, can wait for a suitable opportunity to interject.

Knowing what informational content might help is not the same as knowing when to speak. Group deliberation among human teams is a complex dynamic. The system driving co-agents therefore requires "groupwise etiquette" that feels natural to human teams. This is achieved by coupling triage to inter-contribution timing (ICT) gates that weigh informational urgency against elapsed conversation since the agent last spoke. Immediate, high-urgency needs may enter sooner (someone asks a question); when needs are milder, the co-agent holds back and may gather additional information that can later deepen a line of reasoning or open a productive tangent. Proactive interjections are thus gated by group etiquette as well as informational value.

When conditions warrant, a sourcing stage turns the recent groupwise dialog and any triage notes into web search aimed at several concise, sourced facts—often from complementary angles—and merging them into the knowledge base. Composition then draws from that stored knowledge base and is deliberately interlocutory rather than encyclopedic. The result is a compact message in natural dialog that helps the group deliberate, not a dump of everything known. Attribution is attached so participants know the source of the content.

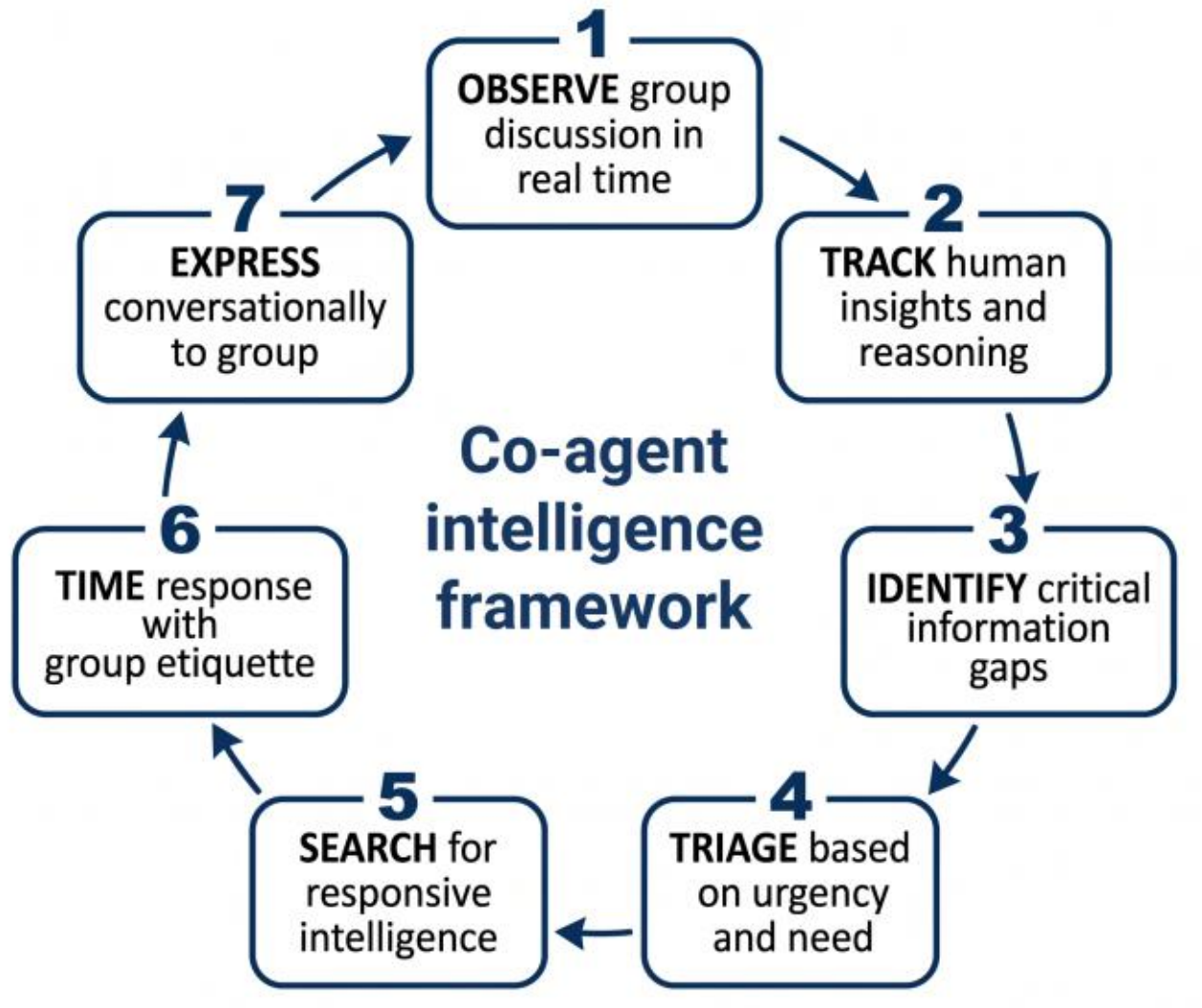

Figure 2: Contributor Agent intelligence framework

The result is a constantly vigilant agentic research engine that proactively and constructively participates in real-time human groupwise discussions, mitigates information barriers within the conversation, equalizes information discrepancies among participants, and increases the rigor of the deliberation without relying on participants to develop their own search queries.

## III. Methods

A study of proactive Contributor Agents was conducted by forecasting 34 Major League Baseball games over 3 weeks of the MLB season in August, collected over five separate "game day" sessions, each with questions spanning two different types of forecasts – (i) predicting team winners and (ii) predicting total number of runs scored. The trials were conducted using a chat-based communication platform called Thinkscape® that can divide any population into small parallel subgroups, and selectively add supportive AI agents to each subgroup that participates with human subgroups in real-time discussion.

Participants were sourced randomly from a panel provider (Prolific) and screened for suitability based on self-identification as a "baseball fan". In each session, that day's participant pool (24-32 participants, varied by game day due to fluctuations in availability from the panel provider) was randomly divided into groups of 4-5. For each question, half the rooms were randomly assigned to the co-agent treatment condition and the other half served as a control, with human-only group conversations. The conversational contributor agent, displayed in each treatment room with the name "(Co)agent Scout", was proactive: able to contribute without an explicit request from a human participant, and had unrestricted scouting abilities with which to (i) identify knowledge gaps in an ongoing discussion, (ii)search the open internet for relevant and constructive information, and (iii) share the sourced information as natural conversational contributions delivered with appropriate groupwise etiquette. It should be noted that Scout could also respond to requests for information from any of the human members of the conversation and/or respond to follow-up questions to any participant regarding provided information.

To ensure equal exposure to forecasting trials with and without the use of co-agents, and to ensure order randomization across the study, the subset of rooms with co-agents were flipped after every question. In addition, the human groups were re-randomized after sets of questions were asked, each set having equal numbers of one treatment and control. Thus, questions had carefully-balanced repeated measures: multiple treatment and control observations were collected for each question, allowing the analysis to control for variation in prediction difficulty between forecasting questions, which helps account for noisy outcomes ("upsets"). *And:* across the pairs of questions, we collected observations of both treatment and control *from each random room*, which helps account for different groups' baseline predictive skills and prior knowledge from one room to another.

During each session, groups collaboratively forecasted a set of MLB games scheduled for play that evening. Half of the trials were A/B Teams forecasts (i.e., which team would win, Team A or Team B, and by how many runs) and half were Over/Under forecasts (i.e., how many total runs would be scored by both teams vs. the Las Vegas consensus line). For each forecasting trial, each small group of participants were asked to discuss and debate an upcoming game and collaboratively converging on a group forecast. The two question formats were:

1. *Who wins and why: Team A (Pitcher A) @Team B (Pitcher B)*

   *Options: [Team A by 3+, Team A by 1-2, Team B by 1-2, Team B by 3+]*

2. *Does this game go Under or Over X runs and why: Team A (Pitcher A) @Team B (Pitcher B)*

   *Options: [Under by 2+, Under by 1, Over by 1, Over by 2+]*

For each of 5 game days, a selection of games were picked to fit some specifications. A/B Teams questions were picked to have some games with large favorites and some games that were toss-ups. Similarly, Over/Under games were selected to have an even spread of games with high, low, and medium total benchmarks, ranging from 7 to 11 runs.

Questions lasted 4 minutes and 30 seconds, with support for each answer being measured in real-time for each participant based on their chat contributions. The analytics engine inside Thinkscape® tracked each participant's real-time sentiment toward each answer option and mapped the group's collective position as a weighted average onto a continuous forecast scale from $-2$ to $+2$ (e.g., $-2$ = full support for "Team A by 3+" or "Under by 2+"; $+2$ = full support for "Team B by 3+" or "Over by 2+"). We refer to this group-level interpolated value as the group's Average Answer, denoted $a_i$.

To evaluate the effect of the co-agent on each group's *collective intelligence*, we operationalize the predictive accuracy of the group as the error margin between their average answer as defined above against the actual outcome, with lower error meaning better forecasts. To limit sensitivity to extreme margins, realized outcomes were mapped to the same five ordinal bins, $c_i^* \in \{-2, -1, 0, +1, +2\}$, according to which multiple-choice option was correct (e.g., a home-team win by two runs → $+1$; home team loss by three runs → $-2$ ). Zeros were used only for Over/Under items that finished exactly on the benchmark total.

Forecast error was partially binned: $a_i$ remained continuous so as to detect small differences in group predictions that would otherwise fall in the same bin, while $c_i^*$ was binned, as the values were discrete so no precision is lost by binning. Squared error was $(a_i - c_i^*)^2$. For mixed-model analysis we used $Y_i = | a_i - c_i^* |$, the square root of squared error, which reduces right-skew to better satisfy the normality and homoscedasticity assumptions of the linear mixed model while letting coefficients be read as marginal changes in absolute forecast error on the $[-2, 2]$ scale.

To measure co-agent effects on group forecasting accuracy, we fit linear mixed-effects models with contributor status as a fixed effect and crossed random intercepts for question $q(i)$ (game-to-game predictability) and session-scoped room $r(i)$ (group baseline knowledge). This structure matches the crossover design: multiple treatment and control rooms forecast each game, and each room contributed observations in both conditions before re-randomization. Models used outcome $Y_i = | a_i - c_i^* |$ (defined above), were estimated by REML, and contributor effects were evaluated with two-sided Wald tests.

Primary analyses fit separate models for A/B Teams and Over/Under questions, reflecting the distinct forecasting tasks in the design:

$$Y_i = \beta_0 + \beta_1 C_i + \alpha_{q(i)} + \omega_{r(i)} + \varepsilon_i$$

fit to each question-type subset, where $C_i = 1$ for co-agent rooms and $C_i = 0$ for human-only control (reference); $\beta_1$ is the co-agent effect within that question type.

A Co-agent × Question Type interaction model was fit on the full sample to assess whether the co-agent effect differed by question type:

$$Y_i = \beta_0 + \beta_1 C_i + \beta_2 O_i + \beta_3 C_i O_i + \alpha_{q(i)} + \omega_{r(i)} + \varepsilon_i,$$

where $O_i = 1$ for Over/Under and $O_i = 0$ for A/B Teams (reference); $\beta_1$ is the co-agent effect on A/B Teams and $\beta_1 + \beta_3$ on Over/Under.

In all models, $i$ indexes think-tank-by-question observations; $q(i)$ and $r(i)$ map observation $i$ to its question and room; $\alpha_{q(i)}$ and $\omega_{r(i)}$ are the corresponding random intercepts; and $\varepsilon_i$ is the residual error.

## IV. Results and Discussion

Table 1 summarizes co-agent effects from separate linear mixed models fit to each question type (outcome: absolute partially binned forecast error, $Y_i = | a_i - c_i^* |$ ). Co-agent groups showed significantly lower error on A/B Teams questions ($\beta_1 = -0.22$, 95% CI $[-0.37, -0.06]$, $p = .006$). This corresponds to a 25% reduction in mean absolute error when proactive scouting co-agents were present in the group forecasting conversation. On Over/Under questions, the estimated effect was not statistically significant ($\beta_1 = 0.11$, 95% CI $[-0.20, 0.42]$ , $p = .481$), indicating no reliable co-agent impact on that question type.

Table 1: Model Effects of Co-agents on Prediction Error

| Question Type | Co-agent Effect ($\beta_1$) | 95% CI | p-value |
|---|---|---|---|
| A/B Teams | -0.218 | [-0.373,-0.062] | .006 |
| Over/Under | +0.110 | [-0.196, 0.416] | .481 |

Table 2: Model Effects of Interaction Model

| Term | Effect | 95% CI | p-value |
|---|---|---|---|
| Co-agent ($\beta_1$) | -0.224 | [-0.470, 0.021] | .073 |
| Question Type ($\beta_2$) | 0.495 | [-0.067, 1.057] | .084 |
| Co-agent x Question Type ($\beta_3$) | 0.33 | [-0.01, 0.68] | .059 |

To assess whether the co-agent effect differed significantly between question types, we fit a Co-agents × Question Type interaction model on the full sample (Table 2). The co-agent effect on A/B Teams (reference level) was directionally negative but did not reach conventional significance in this combined model ($\beta_1 = -0.22, p = .073$). The interaction term was closer but still short of the significance threshold ($\beta_3 = 0.33$, 95% CI $[-0.01, 0.68]$, $p = .059$), consistent with a larger benefit on A/B Teams than on Over/Under. The implied Over/Under effect ($\beta_1 + \beta_3 = 0.11$) matched the stratified Over/Under estimate and was non-significant. Together, these results indicate that co-agent benefits were concentrated in A/B Teams forecasting, aligning with the primary stratified analyses in Table 1. $\beta_2$

**Further Analyses:** Co-agents were designed to participate conversationally in group discussions at rates similar to human team members. That said, the outcome is not deterministic but depends also on the identification of knowledge gaps within each ongoing discussion. The actual contribution rates of the humans and co-agents are shown in Table 3. The humans spoke slightly more, but it was fairly close, confirming that co-agents successfully mimicked human engagement. Additionally, the human contribution rates did not vary in any significant way whether or not a co-agent was present in the group.

Because of the distribution in co-agent groups with 4 to 5 humans and one co-agent, despite the similar contribution rates on a per user basis, co-agents only accounted for 18% of the total conversation.

Table 3: Human vs Co-agent Participation

| Room Type | User Type | Messages per minute per user | Total messages |
|---|---|---|---|
| Co-agent | Co-agent | 0.59 | 310 (18%) |
| Co-agent | Human | 0.66 | 1456 (82%) |
| Human-only | Co-agent | n/a | n/a |
| Human-only | Human | 0.66 | 1448 |

While the proactive co-agents spoke less than the human participants on average, they provided a great deal of context to their fellow participants in the decision making process. Consider Figure 3 depicting a snapshot of conversation among the members of Subgroup 7. As shown, the group is debating whether the San Francisco Giants (Webb pitching) or Boston Red Sox (Gray pitching) will win and by how much. The local co-agent (i) observes the ongoing debate in real-time, (ii) identifies the need for information that could help group choose between these two teams, (iii) proactively searches online for responsive content, (iv) finds information sourced from MLB Reference that provides sound reasoning why San Francisco could be a good pick, and (v) expresses that content as natural dialog within the ongoing groupwise discussion. This all happens autonomously in real-time, meshing seamlessly with the ongoing discourse.

While the group seems skeptical of San Francisco because of their weak pitching, co-agent provided context that Webb has been a good pitcher in recent outings. This caused a slight shift in belief to only slightly favor Boston, and ultimately predict the Red Sox to win by 1 to 2 runs. In this game on August 21, the Red Sox won by 2 runs.

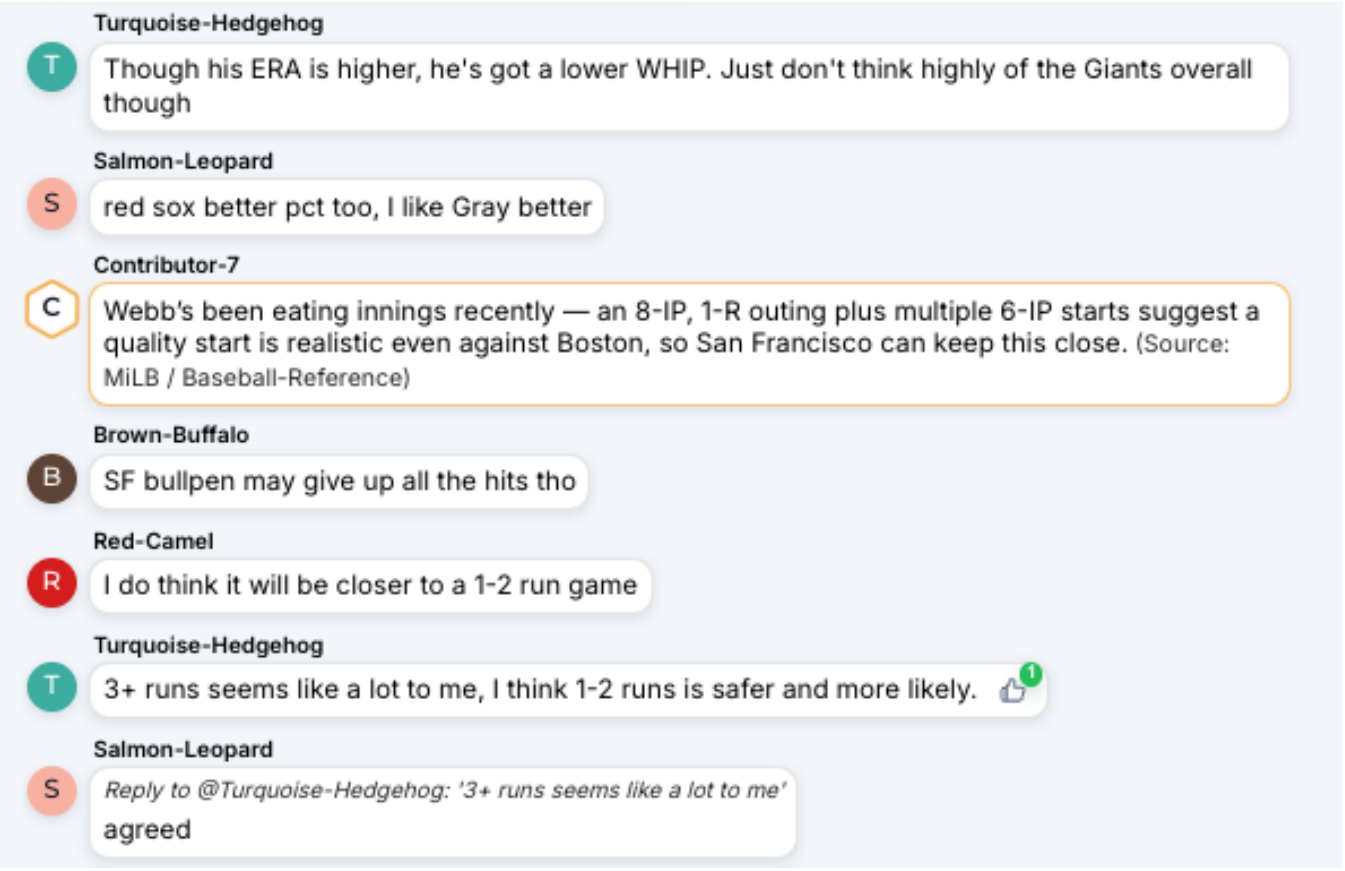

Figure 3: Group Conversation Snapshot with Co-agent

It's important to note that the proactive co-agents in this study were not designed to offer an opinion that could sway the decision for the humans, but rather provide factual evidence and/or reasoning that could help the human members make a more informed and better reasoned decision.

In addition to the mixed-effects model measuring precise error in predictive accuracy, the raw accuracy of picks could also be measured. Table 4 shows the accuracy of the human-only and co-agent individual groups on whether or not the side of the average answer was correct. For games in which the average answer was too close to 0, no pick was recorded, and appears as a tie in the table below. A tie was also recorded for Over/Under forecast games in which the total number of runs scored equaled the benchmark.

Table 4: Raw accuracy of Co-agent and Human-only groups

| **Question Type** | **Room Type** | **Record** (W-L-T) | **Accuracy** (exp. acc.) | **95% CI** |
|---|---|---|---|---|
| A/B Teams | Co-agent Support | 42-13-3 | 76.4% (56.2%) | (63.0%, 86.8%) |
| A/B Teams | Human-only | 33-22-3 | 60.0% (55.1%) | (45.9%, 73.0%) |
| Over/Under | Co-agent Support | 21-30-8 | 41.2% (50.6%) | (27.6%, 55.8%) |
| Over/Under | Human-only | 22-27-10 | 44.9% (52.8%) | (30.7%, 59.8%) |
| All | Co-agent Support | 63-43-11 | 59.4% (53.5%) | (49.5%, 68.9%) |
| All | Human-only | 59-43-13 | 52.9% (53.6%) | (42.8%, 62.8%) |

Both the co-agent and human-only groups were more accurate on the A/B Teams forecasts than the Over/Under forecasts. Despite those forecasts being inherently easier, as shown by the expected accuracy based on Vegas odds, we see that both group types overperformed the expected accuracy on A/B Teams forecasts, and underperformed on Over/Under forecasts. Moreover, co-agent groups were statistically significantly more accurate than the Vegas odds expect ($p=0.002$), and approaching significance in comparison to the human-only groups ($p=0.050$). This is additional strong evidence that co-agents improve the forecasting ability of small groups on questions similar to the A/B Teams questions in this study.

The difference in accuracy between the A/B Teams and Over/Under forecasts is very interesting, as the difference is significant with or without the use of co-agents. Based on the human-only group accuracies, we can confidently infer that the baseline knowledge of the recruited sports fans used in this study was substantially lower for Over/Under questions than the A/B questions. This is likely because A/B sports betting is far more popular than Over/Under betting among members of the general public[11]. This leads us to the hypothesis that co-agents can significantly improve accuracy on forecasts where participants have sufficient baseline knowledge and understanding, but are potentially ineffective when the collaborating participants are inexperienced at the forecasting task and largely guessing.

Following each gameday session, participants were given a short online survey that asked them to reflect on the forecasting experience, with and without the use of proactive contributor agents. A total of 116 surveys were completed in which the following six binary questions were asked:

1. *When present, were the co-agents helpful to the forecasting process?*
2. *When present, did the co-agents help the group arrive at satisfactory predictions?*
3. *Did the co-agents provide useful information?*
4. *Were the co-agent's contributions timely?*
5. *Do you think business teams will want to include co-agents as part of their forecasting processes?*
6. *When do you think your groups made their smartest forecasts today, with or without co-agents?*

Fig. 4 shows the proportion of "Yes" responses to each question and a 95% Confidence Interval for the true proportion of human participants who would agree with the statement under similar conditions. Statistical significance at the 5% level was observed for all six questions, showing that the proactive co-agents were perceived as valuable to discussion. Over 80% of respondents agreed that co-agents provided useful and timely contributions to conversation and helped the group reach better forecasts. Additionally, 89% of respondents indicated that enterprise teams will want to use co-agents in forecasting tasks.

And finally, 77% of respondents believed their predictions were better with the co-agent than without. These subjective results support the quantitative findings above. Specifically, forecasting teams not only improved their predictive accuracy objectively, but also reported statistically significant subjective impressions that co-agents drove smarter forecasts than without.

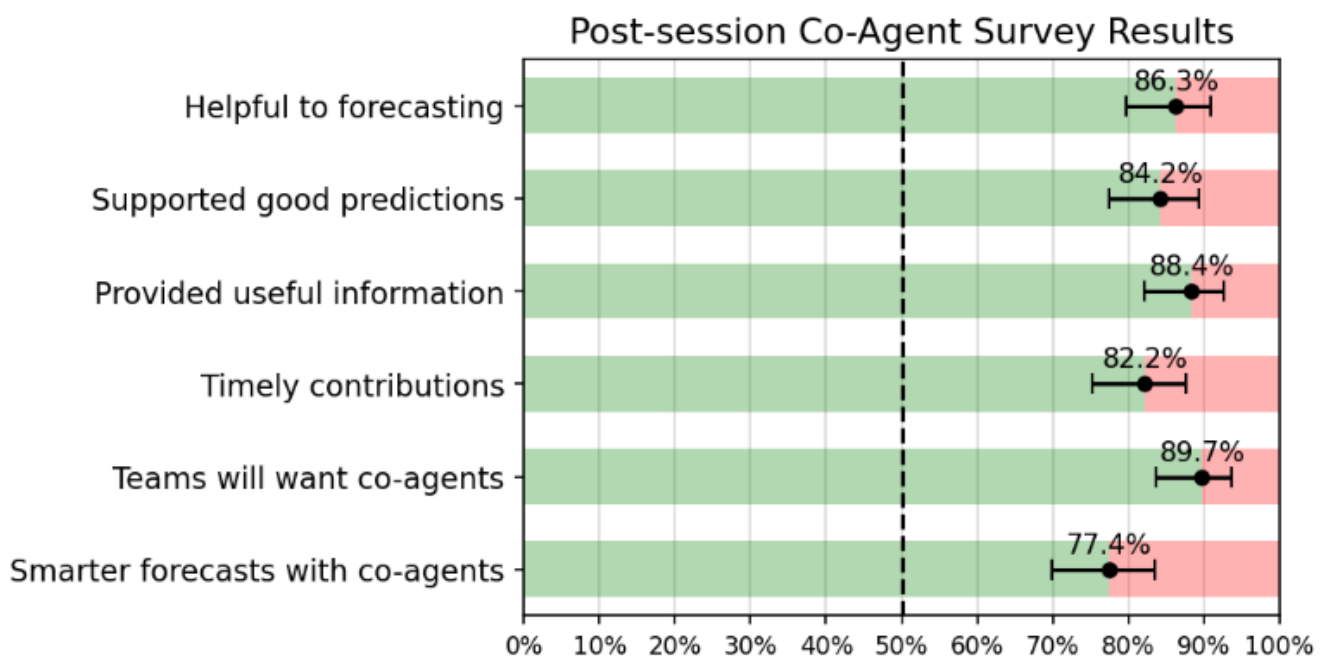

Figure 4: Subjective Impressions of Co-agent Effectiveness

All six questions were found to have statistically significantly higher than 50% agreement (p<0.001), demonstrating that co-agents are seen as beneficial to team forecasting by human participants.

## V. CONCLUSIONS

This study assessed the impact of proactive "scouting" co-agents on groupwise conversational forecasting among small human teams. These conversational agents were designed to: (i) autonomously monitor ongoing discussions among human teams in real-time, (ii) identify informational gaps in the group deliberation, (iii) search for responsive content that is likely to address the gap and thus ensure the team is better informed on issues relevant to the forecast, and (iv) interject the acquired information as natural dialog within the groupwise conversation, expressed within timing that conforms to human groupwise etiquette. The findings demonstrate that proactive co-agents can constructively participate within groupwise deliberations among small human teams without disrupting conversation flow or disturbing group dynamics.

Specifically, the co-agents, when added to human teams of four members, accounted for 18% of the total conversation while human participation rates remained unchanged with or without co-agents. In addition, post-session surveys (N=116) revealed that 82% of human participants reported that the contributions of the co-agents "were timely" within the conversation and 88% of participants reported that the contributions "contained useful information." This suggests that co-agents were perceived as well integrated into the discussion as digital co-workers that provided value.

Regarding predictive performance in the conversational forecasting task, the study yielded the following key outcomes regarding the impact of scouting co-agents:

I. **Significant Improvement in Winner Predictions:** Across 58 groupwise forecasts of the winning team (A/B Teams), the presence of a scouting co-agent significantly increased forecasting accuracy, reducing the mean absolute error by 25% (p=0.006). This demonstrates that proactive scouting agents can significantly assist human forecasting teams during real-time groupwise deliberations.

II. **Increased Raw Accuracy:** For these A/B Team forecasts, raw predictive accuracy jumped from 60.0% in human-only groups to 76.4% in co-agent-supported groups (p=0.050). Furthermore, the co-agent groups performed significantly better than the expected accuracy based on the documented Vegas odds for each MLB game (p=0.002).

III. **Limitations in Over/Under Predictions:** The participation of co-agents did not provide a statistically significant impact forecasting accuracy when tested in Over/Under questions. For these questions, the human participants had low raw accuracy (45%), suggesting they were largely guessing at the outcome. The lack of sufficient baseline skill on the task at hand likely limited the potential impact that co-agents could have on this particular task. For such situations, it may have been desirable for the co-agent to provide higher-level support, suggesting forecasting strategies for Over/Under in addition to providing raw informational assets.

IV. **Human Acceptance of Agentic Team Members:** In addition to finding the contributions of the proactive agents as timely and useful, post-session surveys indicated that 77% of participants felt that the presence of co-agents made their forecasts stronger.

Ultimately, this empirical study provides strong evidence that proactive scouting agents can significantly boost human forecasting accuracy and augment collective intelligence during real-time group meetings. In addition, subjective feedback from the participants indicated overwhelmingly positive reactions to the participation of co-agents, However, this improvement is highly dependent on baseline competence; proactive co-agents are most effective at enhancing human decision-making when both the human participants and the agents already possess a solid foundation of domain knowledge and baseline forecasting skill. With respect to future research, the difference in performance between A/B Teams and Over/Under forecasts raises interesting questions. Future studies of co-agents on different fields of forecasting questions can help uncover the effects of co-agents and where they can help teams make smarter decisions. With co-agents searching the internet in real-time to converse in real-time by providing reasoning, improvements could also be made to better analyze resources used and make effective suggestions.

## ACKNOWLEDGEMENTS

This study was conducted using the **Thinkscape® platform** and the **(Co)agents™** infrastructure from Unanimous AI.